\documentclass[11pt]{article}
\usepackage[latin9]{inputenc}
\usepackage{amsfonts}
\usepackage{amsmath}
\usepackage{amssymb}
\usepackage{indentfirst}
\usepackage{graphicx}
\usepackage[colorlinks]{hyperref}
\usepackage{cite}

\numberwithin{equation}{section}
\begin{document}

\begin{titlepage}
\vspace{3cm}
\baselineskip=24pt

\begin{center}
\textbf{\LARGE{Three-dimensional Poincaré supergravity and $\mathcal{N}$-extended supersymmetric BMS$_{3}$ algebra}}
\par\end{center}{\LARGE \par}

\begin{center}
	\vspace{1cm}
	\textbf{Ricardo Caroca}$^{\ast}$,
	\textbf{Patrick Concha}$^{\ddag}$,
    \textbf{Octavio Fierro}$^{\ast}$,
	\textbf{Evelyn Rodríguez}$^{\dag}$,
	\small
	\\[5mm]
    $^{\ast}$\textit{Departamento de Matemática y Física Aplicadas, }\\
	\textit{ Universidad Católica de la Santísima Concepción, }\\
\textit{ Alonso de Rivera 2850, Concepción, Chile.}
	\\[2mm]
	$^{\ddag}$\textit{Instituto
		de Física, Pontificia Universidad Católica de Valparaíso, }\\
	\textit{ Casilla 4059, Valparaiso-Chile.}
	\\[2mm]

	$^{\dag}$\textit{Departamento de Ciencias, Facultad de Artes Liberales,} \\
	\textit{Universidad Adolfo Ibáñez, Viña del Mar-Chile.} \\[5mm]
	\footnotesize
	\texttt{rcaroca@uscs.cl},
	\texttt{patrick.concha@pucv.cl},
    \texttt{ofierro@uscs.cl},
	\texttt{evelyn.rodriguez@edu.uai.cl},
	\par\end{center}
\vskip 26pt
\begin{abstract}
\noindent
A new approach for obtaining the three-dimensional Chern-Simons supergravity for the Poincaré algebra is presented. The $\mathcal{N}$-extended Poincaré supergravity is obtained by expanding the super Lorentz theory. We extend our procedure to their respective asymptotic symmetries and show that the $\mathcal{N}=(1,2,4)$ super-BMS$_{3}$ appear as expansions of one Virasoro superalgebra. Interestingly, the $\mathcal{N}$-extended super-BMS$_{3}$ obtained here are not only centrally extended but also endowed with internal symmetry. We also show that the $\mathcal{N}$-extended super Poincaré algebras with both central and automorphism generators are finite subalgebras.

\end{abstract}
\end{titlepage}\newpage {}

\section{Introduction}

Three-dimensional (super)gravity theories result of particular interest
since they represent attractive toy models for understanding richer
(super)gravities. Indeed, there are still open issues to solve in higher
dimensions that motivate to study the three-dimensional case. In the last
decades, diverse supergravity models have been presented in three spacetime
dimensions in \cite{DK, Deser, MS, Nieuwenhuizen, RN, NG, BTZs, GTW, GS,
ABRHST, ABRS, BKNTM, NT, ABBOS, BRZ, BDR}. In particular, $\mathcal{N}$%
-extended three-dimensional supergravity without cosmological constant \cite%
{MS} can be expressed as a Chern-Simons (CS) action for the Poincaré
supergroup \cite{AT}. Subsequently, a new class of $\left( p,q\right) $%
-extended Poincaré CS supergravities has been constructed in \cite{HIPT}.
Interestingly, such $\mathcal{N}$-extended flat supergravity with both
central and automorphism charges emerges properly as a vanishing
cosmological constant limit of an $\mathcal{N}$-extended AdS$_{3}$
supergravity \cite{HIPT}.

Recently, there has been a particular interest in the infinite-dimensional
symmetries of asymptotically flat spacetimes at null infinity which were
proposed to be spanned by the BMS algebra originally discovered more than a
half century ago \cite{BBM, Sachs}. In three dimensions, it has been shown
in \cite{ABS, BC, BT1} that the asymptotically flat symmetry is described by
the BMS$_{3}$ algebra. Such infinite-dimensional algebra results to be
isomorphic to the Galilean conformal algebra in two dimensions \cite{Bagchi}%
. Interestingly, the BMS$_{3}$ algebra appears as a flat limit of the
conformal algebra which describe the asymptotic symmetries of
three-dimensional gravity \cite{BH}. More recently, it was shown in \cite%
{CCRS} that the BMS$_{3}$ algebra can be alternatively derived as an
algebraic expansion of the Virasoro one. The derivation of the BMS$_{3}$ symmetry using an algebraic operation has also been considered in \cite{KRR}. Further extensions and deformations
of the BMS$_{3}~$algebra have been recently studied in \cite{GMPT, ABFGR, GP, MPTT, FMT1,
BJMN, DR, SA, CMMRSV, PSSJ, CMRSV}.

At the supersymmetric level, a minimal supersymmetric extension of BMS$_{3}$
appears as the asymptotic symmetry of $\ $a three-dimensional $\mathcal{N}%
=1$ supergravity for suitable boundary conditions \cite{BDMT}. Such
superalgebra turns out to be isomorphic to the Galilean superconformal
algebra \cite{BM, M}. The supersymmetric extension to $\mathcal{N}=2$ \cite%
{LM, FMT}, $\mathcal{N}=4$ \cite{BLN} and $\mathcal{N}=8$ \cite{BBLN}\ of
the BMS$_{3}$ has been subsequently explored by an asymptotic symmetry
analysis. Remarkably, the $\mathcal{N}$-extended super-BMS$_{3}$ can be
obtained by an appropriate contraction of the $\mathcal{N}$-extended
superconformal algebras \cite{BJLMN}.

In this paper, we present a novel approach to obtain the $\mathcal{N}$%
-extended super-BMS$_{3}$ algebra by considering the semigroup expansion
method ($S$-expansion) \cite{Sexp}. The algebraic expansion methods \cite%
{HS, AIPV, Sexp} have been particularly useful in the context of
(super)gravity theories \cite{EHTZ, AIPV1, IRS, IPRS, IMPRS, AI, DFIMRSV,
CPRS1, GRSS, CPRS2, CPRS3, CR2, CRS, CDIMR, CMR, PR1, PR, CCFRS, PR2}. In
three dimensions, the $S$-expansion procedure has not only allowed to
reproduce known (super)gravity theories \cite{CFR} but also to obtain novel
(super)gravity actions \cite{FISV, CFRS, CPR}. Here, we will show first that
the three-dimensional $\mathcal{N}$-extended Poincaré CS supergravity theory
can be alternatively derived from a super-Lorentz CS theory using a
particular semigroup. Interestingly, such procedure can be extended to
infinite-dimensional algebras allowing us to reproduce the $\mathcal{N}$%
-extended super-BMS$_{3}$ algebra from the super Virasoro algebra using the
same finite semigroup. Let us note that the procedure considered here,
unlike the contraction, requires only one $\mathcal{N}$-extended super
Virasoro algebra instead of two copies. In particular, the $\mathcal{N}>1$ super-BMS$%
_{3}$ algebras obtained here are non only centrally extended but also
contain internal symmetry algebra. Thus, the finite subalgebra corresponds
to the $\mathcal{N}$-extended Poincaré superalgebra endowed with
automorphism generators. It is important to point out that the $\mathcal{N}=2$ super-BMS$_{3}$ presented here can be easily recover from the $\mathcal{N}=4$ one presented in \cite{BJLMN} after setting some fermionic generators to zero. The results obtained here can be seen as a
supersymmetric generalizations of those presented in \cite{CCRS} and could
be useful to derive new infinite-dimensional superalgebras.

The paper is organized as follows: In Section 2, we apply the $S$-expansion
method to obtain the $\mathcal{N}$-extended Poincaré supergravity from an $%
\mathcal{N}$-extended super-Lorentz theory in three spacetime dimensions. A
brief introduction to the $\mathcal{N}$-extended super-Lorentz CS theory is
also presented. In Section 3, we extend our procedure to
infinite-dimensional superalgebras. In particular, the \thinspace $\mathcal{N%
}$-extended super-BMS$_{3}$ are obtained from an $\mathcal{N}$-extended
super-Virasoro algebra for $\mathcal{N}=1,2$ and $4$. We show that the $%
\mathcal{N}$-extended super-Poincaré algebra is a finite subalgebra of the
supersymmetric extension of the BMS$_{3}$ algebra. We also discuss the
presence of internal symmetry algebra in the $\mathcal{N}$-extended super-BMS%
$_{3}$ obtained here for $\mathcal{N}>1$. In Section 4, we conclude with
some comments about possible developments and extensions of our results.

\section{$\mathcal{N}$-extended Poincaré supergravity and super-Lorentz
theory in three spacetime dimensions}

The possibility of having a well defined three-dimensional AdS CS gravity
action from a Lorentz action using the $S$-expansion procedure has been
considered in \cite{IPRS}. Here, we extend such result to the $\mathcal{N}$%
-extended Poincaré CS supergravity in three spacetime dimensions. In
particular, the three-dimensional $\mathcal{N}$-extended Poincaré
superalgebra can be obtained from a supersymmetric extension of the Lorentz
algebra using the $S$-expansion. Such method allows us to obtain the
non-vanishing components of the invariant tensor of the super Poincaré which
are essential to construct a CS action. However, it is important to point
out that the procedure presented here can be applied only in three spacetime
dimensions. This particular accident comes from the fact that the expanded
Lorentz generators can be interpreted as translational generators.

\subsection{$\mathcal{N}$-extended super-Lorentz theory}

A supersymmetric extension of the Lorentz algebra in three spacetime
dimensions is generated by the bosonic set $\left\{ M_{a},\text{ }\tilde{T}%
^{ij}\right\} $ and Majorana fermionic generators $\left\{ \tilde{Q}%
^{i}\right\} $ with $i,j=1,\dots ,\mathcal{N}$. The non-vanishing
(anti-)commutation relations of an $\mathcal{N}$-extended super-Lorentz
algebra read%
\begin{eqnarray}
\left[ M_{a},M_{b}\right]  &=&\epsilon _{abc}M^{c}\,,  \notag \\
\left[ M_{a},\tilde{Q}_{\alpha }^{i}\right]  &=&\frac{1}{2}\,\left( \Gamma
_{a}\right) _{\text{ }\alpha }^{\beta }\tilde{Q}_{\beta }^{i}\,,\text{ \ \ }%
\left[ \tilde{T}^{ij},\tilde{Q}_{\alpha }^{k}\right] =\left( \delta ^{jk}%
\tilde{Q}_{\alpha }^{i}-\delta ^{ik}\tilde{Q}_{\alpha }^{j}\right) \,,
\notag \\
\left[ \tilde{T}^{ij},\tilde{T}^{kl}\right]  &=&\delta ^{jk}\tilde{T}%
^{il}-\delta ^{ik}\tilde{T}^{jl}-\delta ^{jl}\tilde{T}^{ik}+\delta ^{il}%
\tilde{T}^{jk}\,, \\
\left\{ \tilde{Q}_{\alpha }^{i},\tilde{Q}_{\beta }^{j}\right\}  &=&-\frac{1}{%
2}\,\delta ^{ij}\left( C\Gamma ^{a}\right) _{\alpha \beta }M_{a}+C_{\alpha
\beta }\tilde{T}^{ij}\,,  \notag
\end{eqnarray}%
where $a,b,\dots =0,1,2$ are Lorentz indices, $\Gamma _{a}$ denote the Dirac
matrices and $C$ represents the charge conjugation matrix,%
\begin{equation}
C_{\alpha \beta }=C^{\alpha \beta }=%
\begin{pmatrix}
0 & -1 \\
1 & 0%
\end{pmatrix}%
\,.
\end{equation}%
In particular, the Dirac Matrices can be written in terms of the Pauli
matrices $\sigma _{i}$ as%
\begin{equation}
\begin{tabular}{lllll}
$\Gamma _{0}=\frac{1}{\sqrt{2}}\left( \sigma _{1}+i\sigma _{2}\right) $ & ,
& $\Gamma _{1}=\frac{1}{\sqrt{2}}\left( \sigma _{1}-i\sigma _{2}\right) $ & ,
& $\Gamma _{2}=\sigma _{3}\,,$%
\end{tabular}%
\,
\end{equation}%
with%
\begin{equation}
\begin{tabular}{lllll}
$\sigma _{1}=%
\begin{pmatrix}
0 & 1 \\
1 & 0%
\end{pmatrix}%
$ & , & $\sigma _{2}=%
\begin{pmatrix}
0 & -i \\
i & 0%
\end{pmatrix}%
$ & , & $\sigma _{3}=%
\begin{pmatrix}
1 & 0 \\
0 & -1%
\end{pmatrix}%
\,.$%
\end{tabular}%
\,
\end{equation}%
Here $\tilde{T}^{ij}=-\tilde{T}^{ji}$ are internal symmetry generators with $%
i=1,\dots ,\mathcal{N}$.

Let us note that the Lorentz superalgebra introduced in \cite{LuNo}
corresponds to the $\mathcal{N}=1$ case and it is spanned by the set of
generators $\left\{ M_{a},\tilde{Q}_{\alpha }\right\} $. On the other hand,
the $\mathcal{N}=2$ Lorentz superalgebra implies the introduction of the $%
\mathfrak{so}\left( 2\right) $ automorphism algebra through the $\tilde{T}$
generator.

A CS action can be constructed from the Lorentz supergroup which has the
following non-vanishing components of the invariant tensor,%
\begin{eqnarray}
\left\langle M_{a}M_{b}\right\rangle  &=&\eta _{ab}\,,  \notag \\
\left\langle \tilde{Q}_{\alpha }^{i}\tilde{Q}_{\beta }^{j}\right\rangle
&=&C_{\alpha \beta }\delta ^{ij}\,,  \label{it} \\
\left\langle \tilde{T}^{ij}\tilde{T}^{kl}\right\rangle  &=&\delta
^{ik}\delta ^{lj}-\delta ^{il}\delta ^{kj}\,,  \notag
\end{eqnarray}%
where $\eta _{ab}$ is the Minkowski metric. On the other hand, let us
consider the gauge connection one-form $A$,%
\begin{equation}
A=\omega ^{a}M_{a}+\bar{\psi}^{i}\tilde{Q}^{i}+\frac{1}{2}\,A^{ij}\tilde{T}%
_{ij}\,\,,  \label{l1f}
\end{equation}%
where the coefficients in front of the generators correspond to the gauge
potential one-forms. In particular, the gauge fields $\omega ^{a}$, $\psi $
are the spin connection and the Majorana spinor field, respectively. The
field strength two-form $F=dA+\frac{1}{2}\left[ A,A\right] $ reads%
\begin{equation}
F=F^{a}M_{a}+\nabla \bar{\psi}^{i}\tilde{Q}^{i}+F^{ij}\tilde{T}^{ij}\,,
\label{FS}
\end{equation}%
with%
\begin{eqnarray}
F^{a} &=&d\omega ^{a}+\frac{1}{2}\epsilon ^{abc}\omega _{b}\omega _{c}+\frac{%
1}{4}\,\bar{\psi}^{i}\Gamma ^{a}\psi ^{i}\,,  \notag \\
F^{ij} &=&dA^{ij}+A^{ik}A^{kj}+\bar{\psi}^{i}\psi ^{j}\,.
\end{eqnarray}%
Here, the covariant derivative acting on spinors reads%
\begin{equation}
\nabla \psi ^{i}=d\psi ^{i}+\frac{1}{2}\,\omega ^{a}\Gamma _{a}\psi
^{i}+A^{ij}\psi ^{j}\,.
\end{equation}%
The CS action%
\begin{equation}
I_{CS}=\frac{k}{4\pi }\int \left\langle AdA+\frac{2}{3}\,A^{3}\right\rangle
\,,
\end{equation}%
can be written, using the invariant tensor (\ref{it}) and the connection
one-form (\ref{l1f}) as%
\begin{equation}
I_{CS}^{\left( 2+1\right) }=\frac{k}{4\pi }\,\int \,\omega ^{a}d\omega _{a}+%
\frac{1}{3}\,\epsilon _{abc}\omega ^{a}\omega ^{b}\omega ^{c}-\frac{1}{2}\,%
\mathcal{G}(A^{ij})-\bar{\psi}^{i}\nabla \psi ^{i}\,,  \label{sLa}
\end{equation}%
where the three-form $\mathcal{G}$ is defined as%
\begin{equation}
\mathcal{G}\left( A^{ij}\right) =A^{ij}dA^{ji}+\frac{2}{3}%
A^{ik}A^{km}A^{mi}\,.
\end{equation}%
The field equations are given by the vanishing of the coefficients appearing
in the field strength (\ref{FS})%
\begin{equation}
F^{a}=0\,,\text{\qquad }\nabla \psi ^{i}=0\,,\qquad F^{ij}=0\,.
\end{equation}%
Note that the Lagrangian (\ref{sLa}) contains the exotic Lagrangian, also
known as Lorentz Lagrangian, $L_{exotic}=\omega ^{a}d\omega _{a}+\frac{1}{3}%
\,\epsilon _{abc}\omega ^{a}\omega ^{b}\omega ^{c}$. The above
three-dimensional action describes the coupling of Majorana spinor field to
the exotic term and to an $SO\left( \mathcal{N}\right) $ CS term. Such
action is invariant, up to boundary terms, under the gauge transformation $%
\delta A=D\lambda =d\lambda +\left[ A,\lambda \right] \,$. In particular,
the non-zero supersymmetry transformation laws are given by%
\begin{eqnarray}
\delta \omega ^{a} &=&\frac{1}{2}\,\bar{\epsilon}^{i}\Gamma ^{a}\psi ^{i}\,,
\notag \\
\delta \psi ^{i} &=&\nabla \epsilon ^{i}\,, \\
\delta A^{ij} &=&-2\bar{\psi}^{\left[ i\right. }\epsilon ^{\left. j\right] }%
\text{\thinspace },  \notag
\end{eqnarray}%
where one can see that the $SO\left( \mathcal{N}\right) $ automorphism gauge
fields are present. It is no a surprise that the action (\ref{sLa}) is
invariant under the super-Lorentz group since it is built from the gauge
connection one-form $A$ as a CS action.

\subsection{$\mathcal{N}$-extended Poincaré supergravity theory from
super-Lorentz}

The $\mathcal{N}$-extended CS Poincaré supergravity theory in three
spacetime dimensions in presence of both central and automorphism charges,
has been carefully studied in \cite{HIPT}. Although the $\mathcal{N}$%
-extended Poincaré supergravity has been discussed much earlier by diverse
authors \cite{MS, AT}, the $\mathcal{N}=p+q$ super Poincaré one presented in
\cite{HIPT} has interesting advantages. In particular, the $\left(
p,q\right) $-extended Poincaré supergravity appears as a Poincaré limit of
the corresponding $\left( p,q\right) $-extended AdS supergravity theories
\cite{HIPT, CFR}. Additionally, the $\left( 1,1\right) $ and $\left(
2,0\right) $ super Poincaré theories possess an off-shell superfield
formulation.

In this section, following a similar procedure used in \cite{CCRS}, we
present a novel method to recover the $\mathcal{N}$-extended Poincaré
supergravity theory. We show that an algebraic expansion mechanism can be
applied to the super-Lorentz theory introduced in the previous section. Such
method allows not only to recover the complete set of (anti-)commutation
relations of the $\mathcal{N}$-extended Poincaré superalgebra (including
both central and automorphism charges) but also the complete CS supergravity
action. As we shall see in the next section, the procedure used here can be
generalized at the asymptotic level.

The super-Lorentz algebra can be decomposed in subspaces as%
\begin{equation}
s\mathcal{L}=V_{0}\oplus V_{1}\,,
\end{equation}%
where $V_{0}$ is the bosonic subspace spanned by the Lorentz generator $M_{a}
$ and the automorphism generators $\tilde{T}^{ij}$. On the other hand, $%
V_{1}$ is the fermionic subspace. Such subspaces satisfy a graded Lie
algebra,
\begin{eqnarray}
\left[ V_{0},V_{0}\right]  &\subset &V_{0}\,,  \notag \\
\left[ V_{0},V_{1}\right]  &\subset &V_{1}\,, \\
\left[ V_{1},V_{1}\right]  &\subset &V_{0}\,.  \notag
\end{eqnarray}%
Interestingly, the $\mathcal{N}$-extended super Poincaré structure can be
recovered from it using the $S$-expansion method \cite{Sexp}. However, it is
necessary that the pertinent semigroup $S$ possess a particular
decomposition $S=S_{0}\cup S_{1}$ which has to behave as the subspaces of
the super-Lorentz algebra. An abelian semigroup with the desired behavior is
$S_{E}^{\left( 2\right) }=\left\{ \lambda _{0},\lambda _{1},\lambda
_{2},\lambda _{3}\right\} $ whose elements satisfy%
\begin{equation}
\begin{tabular}{l|llll}
$\lambda _{3}$ & $\lambda _{3}$ & $\lambda _{3}$ & $\lambda _{3}$ & $\lambda
_{3}$ \\
$\lambda _{2}$ & $\lambda _{2}$ & $\lambda _{3}$ & $\lambda _{3}$ & $\lambda
_{3}$ \\
$\lambda _{1}$ & $\lambda _{1}$ & $\lambda _{2}$ & $\lambda _{3}$ & $\lambda
_{3}$ \\
$\lambda _{0}$ & $\lambda _{0}$ & $\lambda _{1}$ & $\lambda _{2}$ & $\lambda
_{3}$ \\ \hline
& $\lambda _{0}$ & $\lambda _{1}$ & $\lambda _{2}$ & $\lambda _{3}$%
\end{tabular}%
\,  \label{ml}
\end{equation}%
where $\lambda _{3}=0_{s}$ is the zero element of the semigroup. In
particular, the following subset decomposition $S_{E}^{\left( 2\right)
}=S_{0}\cup S_{1}$, with%
\begin{eqnarray}
S_{0} &=&\left\{ \lambda _{0},\lambda _{2},\lambda _{3}\right\} \,,  \notag
\label{sd1} \\
S_{1} &=&\left\{ \lambda _{1},\lambda _{3}\right\} \,,  \label{sd2}
\end{eqnarray}%
is said to be resonant since it satisfies the same subspace's structure,%
\begin{eqnarray}
S_{0}\cdot S_{0} &\subset &S_{0}\,,  \notag \\
S_{0}\cdot S_{1} &\subset &S_{1}\,, \\
S_{1}\cdot S_{1} &\subset &S_{0}\,.  \notag
\end{eqnarray}%
Following the definitions of \cite{Sexp}, after extracting a resonant
subalgebra of $S_{E}^{\left( 2\right) }\times s\mathcal{L}$ and applying its
$0_{s}$-reduction, one finds an expanded superalgebra whose generators $%
\left\{ J_{a},P_{a},T^{ij},Z^{ij},Q^{i}\right\} $ are related to the
super-Lorentz one as%
\begin{eqnarray}
J_{a} &=&\lambda _{0}M_{a}\,,  \notag \\
P_{a} &=&\lambda _{2}M_{a}\,,  \notag \\
T^{ij} &=&\lambda _{0}\tilde{T}^{ij}\,, \\
Z^{ij} &=&\lambda _{2}\tilde{T}^{ij}\,,  \notag \\
Q^{i} &=&\lambda _{1}\tilde{Q}^{i}\,.  \notag
\end{eqnarray}%
Using the multiplication law of the semigroup, one can see that the
generators of the expanded superalgebra satisfy the following non-vanishing
(anti-)commutation relations%
\begin{eqnarray}
\left[ J_{a},J_{b}\right]  &=&\epsilon _{abc}J^{c}\,,  \notag \\
\left[ J_{a},P_{b}\right]  &=&\epsilon _{abc}P^{c}\,,  \notag \\
\left[ J_{a},Q_{\alpha }^{i}\right]  &=&\frac{1}{2}\,\left( \Gamma
_{a}\right) _{\text{ }\alpha }^{\beta }Q_{\beta }^{i}\,,\text{ \ \ \
}  \label{sp1} \\
\left\{ Q_{\alpha }^{i},Q_{\beta }^{j}\right\}  &=&-\frac{1}{2}\,\delta
^{ij}\left( C\Gamma ^{a}\right) _{\alpha \beta }P_{a}+C_{\alpha \beta
}Z^{ij}\,,  \notag
\end{eqnarray}%
\begin{eqnarray}
\left[ T^{ij},Q_{\alpha }^{k}\right]  &=&\left( \delta ^{jk}Q_{\alpha
}^{i}-\delta ^{ik}Q_{\alpha }^{j}\right) \,,  \notag \\
\left[ T^{ij},T^{kl}\right]  &=&\delta ^{jk}T^{il}-\delta ^{ik}T^{jl}-\delta
^{jl}T^{ik}+\delta ^{il}T^{jk}\,,  \label{sp2} \\
\left[ T^{ij},Z^{kl}\right]  &=&\delta ^{jk}Z^{il}-\delta ^{ik}Z^{jl}-\delta
^{jl}Z^{ik}+\delta ^{il}Z^{jk}\,.  \notag
\end{eqnarray}%
The (anti-)commutation relations given by (\ref{sp1}) correspond to the
central extension of the $\mathcal{N}$-extended Poincaré superalgebra. \
Interestingly, the $S$-expansion procedure also provides with the
automorphism charges which satisfy (\ref{sp2}). In particular, unlike the
contraction method in which some (anti-)commutators vanish, the expansion
provides us with a bigger algebra with a new set of (anti-)commutators.
Here, the set of (anti-)commutation relations is known as the complete $%
\mathcal{N}$-extended Poincaré superalgebra \cite{HIPT}.

As was mentioned in \cite{HIPT}, the superalgebra (\ref{sp1}) does not admit
a non-degenerate invariant inner product which prevent the formulation of a
CS supergravity action. It is the presence of the automorphism group which
allows a CS supergravity formulation. Note that the the central charges $%
Z^{ij}$ do not need to be invariant under the automorphism algebra.

Remarkably, the $S$-expansion does not limit only to the obtention of the
super-Poincaré generators but also provides us with the non-vanishing
components of the invariant tensor of the $\mathcal{N}$-extended Poincaré
supergravity action. In fact, following the definitions of \cite{Sexp}, the
Poincaré invariant tensor is given in term of the super-Lorentz one:%
\begin{eqnarray}
\left\langle J_{a}J_{b}\right\rangle &=&\mu _{0}\left\langle
M_{a}M_{b}\right\rangle =\mu _{0}\eta _{ab}\,,  \notag \\
\left\langle J_{a}P_{b}\right\rangle &=&\mu _{2}\left\langle
M_{a}M_{b}\right\rangle =\mu _{2}\eta _{ab}\,,  \notag \\
\left\langle Q_{\alpha }^{i}Q_{\beta }^{j}\right\rangle &=&\mu
_{2}\left\langle \tilde{Q}_{\alpha }^{i}\tilde{Q}_{\beta }^{j}\right\rangle
=\mu _{2}C_{\alpha \beta }\delta ^{ij}\,,  \label{itSP} \\
\left\langle T^{ij}T^{kl}\right\rangle &=&\mu _{0}\left\langle \tilde{T}^{ij}%
\tilde{T}^{kl}\right\rangle =\mu _{0}\left( \delta ^{ik}\delta ^{lj}-\delta
^{il}\delta ^{kj}\right) \,,  \notag \\
\left\langle Z^{ij}T^{kl}\right\rangle &=&\mu _{2}\left\langle \tilde{T}^{ij}%
\tilde{T}^{kl}\right\rangle =\mu _{2}\left( \delta ^{ik}\delta ^{lj}-\delta
^{il}\delta ^{kj}\right) \,,  \notag
\end{eqnarray}%
where $\mu _{0}$ and $\mu _{2}$ are arbitrary constants.

The connection one-form reads
\begin{equation}
A=\omega ^{a}J_{a}+e^{a}P_{a}+\bar{\psi}^{i}Q^{i}+\frac{1}{2}%
\,A^{ij}T_{ij}\,+\frac{1}{2}\,C^{ij}Z_{ij}\,,  \label{1fP}
\end{equation}%
where the coefficients in front of the generators are the gauge potential
one-forms. Let us note that the vielbein $e^{a}$, which was not present in
the super Lorentz theory, appears naturally from the $S$-expansion
procedure. Furthermore, the methodology provides us with central gauge fields $%
C^{ij}$ in addition to the automorphism gauge fields $A^{ij}$.

The curvature two-form is given by%
\begin{equation}
F=R^{a}J_{a}+T^{a}P_{a}+\nabla \bar{\psi}^{i}Q^{i}+\frac{1}{2}\,F^{ij}T_{ij}+%
\frac{1}{2}\,G^{ij}Z_{ij}\,,
\end{equation}%
where%
\begin{eqnarray}
R^{a} &=&d\omega ^{a}+\frac{1}{2}\epsilon ^{abc}\omega _{b}\omega _{c}\,,
\notag \\
T^{a} &=&de^{a}+\epsilon ^{abc}\omega _{b}e_{c}+\frac{1}{4}\bar{\psi}%
^{i}\Gamma ^{a}\psi ^{i}\,,
\end{eqnarray}%
are the Lorentz curvature and supertorsion curvature, respectively. On the
other hand,
\begin{eqnarray}
\nabla \psi ^{i} &=&d\psi ^{i}+\frac{1}{2}\,\omega ^{a}\Gamma _{a}\psi
^{i}+A^{ij}\psi ^{j}\,,  \notag \\
F^{ij} &=&dA^{ij}+A^{ik}A^{kj}\,, \\
G^{ij} &=&dC^{ij}+C^{ik}A^{kj}+A^{ik}C^{kj}-\bar{\psi}^{i}\psi ^{j}\,.
\notag
\end{eqnarray}%
Let us note that the two-form curvature related to the automorphism gauge
fields has no longer spinor fields as in the super-Lorentz case.

The CS supergravity action can be written considering the non-vanishing
components of the invariant tensor (\ref{itSP}) and the gauge connection
one-form (\ref{1fP}),%
\begin{eqnarray}
I_{CS}^{\left( 2+1\right) } &=&\frac{k}{4\pi }\int \mu _{0}\left[ \,\omega
^{a}d\omega _{a}+\frac{1}{3}\,\epsilon _{abc}\omega ^{a}\omega ^{b}\omega
^{c}-\frac{1}{2}\,\mathcal{G}(A^{ij})\right]  \notag \\
&&+\mu _{1}\left[ 2e^{a}R_{a}-\bar{\psi}^{i}\nabla \psi ^{i}-\frac{1}{2}%
\,C^{ij}F^{ji}\right] \,,  \label{csp}
\end{eqnarray}%
with%
\begin{equation}
\mathcal{G}\left( A^{ij}\right) =A^{ij}dA^{ji}+\frac{2}{3}%
A^{ik}A^{km}A^{mi}\,.
\end{equation}%
$\,$

Notice that the term proportional to $\mu _{0}$ contains the exotic
Lagrangian plus contributions coming from the automorphism gauge fields.
Unlike the $\mathcal{N}$-extended $AdS$ supergravity theories, the gravitini
do not appear in the exotic sector.

By construction, the CS action (\ref{csp}) is invariant under the gauge
transformation $\delta A=D\lambda =d\lambda +\left[ A,\lambda \right] \,$.
In particular, the action is invariant under the following local
supersymmetry transformation laws%
\begin{eqnarray}
\delta \omega ^{ab} &=&0\,,  \notag \\
\delta e^{a} &=&\frac{1}{2}\,\bar{\epsilon}^{i}\Gamma ^{a}\psi ^{i}\,,
\notag \\
\delta \psi ^{i} &=&\nabla \epsilon ^{i}\,, \\
\delta A^{ij} &=&0\,,  \notag \\
\delta C^{ij} &=&-2\bar{\psi}^{\left[ i\right. }\epsilon ^{\left. j\right] }%
\text{\thinspace }.  \notag
\end{eqnarray}

The new procedure introduced here, allowing us to recover the $\mathcal{N}$%
-extended Poincaré supergravity theory, can be generalized to obtain the
asymptotic symmetry of the $\mathcal{N}$-extended Poincaré supergravity. In
the next section, we show first how to obtain the $\mathcal{N}=1$ super-BMS$%
_{3}$ algebra and then we naturally extend our study to the $\mathcal{N}$%
-extended case.

\section{$\mathcal{N}$-extended super-BMS$_{3}$ algebra from $\mathcal{N}$%
-extended super-Virasoro algebra}

It is well known that the BMS$_{3}$ symmetry emerges as a suitable
contraction of the asymptotic symmetry of the AdS gravity, which is given by
two copies of the Virasoro algebra. Analogously, the supersymmetric
extension of the BMS$_{3}$ algebra comes by performing an Inönü-Wigner
contraction to appropriate superconformal algebras \cite{BLN, BJLMN}.

In this section, we show a new way to obtain the $\mathcal{N}$-extended
super-BMS$_{3}$ algebra from only one copy of the $\mathcal{N}$-extended
super Virasoro algebra. This procedure corresponds to a supersymmetric
extension of the results presented in \cite{CCRS}. In particular,
considering the same semigroup of the previous section, we obtain the $%
\mathcal{N}=1,2$ and $4$ super-BMS$_{3}$ algebra whose finite subalgebra is
the $\mathcal{N}$-extended super-Poincaré one.

\subsection{$\mathcal{N}=1$ super-BMS$_{3}$ algebra\qquad}

The starting point of our construction is the super-Virasoro algebra, which
we will denote as $\mathfrak{svir}$, whose (anti)-commutation relations are
given by%
\begin{eqnarray}
\left[ \ell _{m},\ell _{n}\right] &=&\left( m-n\right) \ell _{m+n}+\frac{c}{%
12}\,m\left( m^{2}-1\right) \delta _{m+n,0}\,,  \notag \\
\left[ \ell _{m},\mathcal{Q}_{r}\right] &=&\left( \frac{m}{2}-r\right)
\mathcal{Q}_{m+r}\,, \\
\left\{ \mathcal{Q}_{r},\mathcal{Q}_{s}\right\} &=&\ell _{r+s}+\frac{c}{6}%
\,\left( r^{2}-\frac{1}{4}\right) \delta _{r+s,0}\,.  \notag
\end{eqnarray}%
Let us note that the super-Lorentz algebra corresponds to a finite
subalgebra of the super-Virasoro one. Indeed, the three-dimensional
super-Lorentz algebra is spanned by the generators $\ell _{0},\ell _{1},\ell
_{-1},\mathcal{Q}_{\pm \frac{1}{2}}$ which are related to the super-Lorentz
generators through the following change of basis:%
\begin{equation}
\begin{tabular}{lllll}
$\ell _{-1}=-\sqrt{2}M_{0}$ & , & $\ell _{1}=\sqrt{2}M_{1}$ & , & $\ell
_{0}=M_{2}\,,$ \\
$\mathcal{Q}_{-\frac{1}{2}}=\sqrt{2}\tilde{Q}_{+}$ & , & $\mathcal{Q}_{\frac{%
1}{2}}=\sqrt{2}\tilde{Q}_{-}\,.$ &  &
\end{tabular}%
\end{equation}

Let us consider now $S_{E}^{\left( 2\right) }=\left\{ \lambda _{0},\lambda
_{1},\lambda _{2},\lambda _{3}\right\} $ whose elements satisfy (\ref{ml}).
After extracting a resonant subalgebra of $S_{E}^{\left( 2\right) }\times
\mathfrak{svir}$ and performing a $0_{s}$-reduction, a new set of generators
is obtained. In fact, the expanded algebra consists of the set of generators $\left\{
\mathcal{J}_{m},\mathcal{P}_{m},\mathcal{G}_{r},c_{1},c_{2}\right\} $ which
are related to the super-Virasoro ones through the semigroup elements in
the following way:%
\begin{equation}
\begin{tabular}{ll}
$\mathcal{J}_{m}=\lambda _{0}\ell _{m}\,,$ & $c_{1}=\lambda _{0}c\,,$ \\
$\mathcal{P}_{m}=\lambda _{2}\ell _{m}\,,$ & $c_{2}=\lambda _{2}c\,,$ \\
$\mathcal{G}_{r}=\lambda _{1}\mathcal{Q}_{r}\,.$ &
\end{tabular}%
\end{equation}%
Using the (anti-)commutators of the super-Virasoro algebra together with the
multiplication law of the semigroup (\ref{ml}), one find that the
non-vanishing (anti-)commutation relations of the expanded algebra are%
\begin{eqnarray}
\left[ \mathcal{J}_{m},\mathcal{J}_{n}\right] &=&\left( m-n\right) \mathcal{J%
}_{m+n}+\frac{c_{1}}{12}\,m\left( m^{2}-1\right) \delta _{m+n,0}\,,  \notag
\\
\left[ \mathcal{J}_{m},\mathcal{P}_{n}\right] &=&\left( m-n\right) \mathcal{P%
}_{m+n}+\frac{c_{2}}{12}\,m\left( m^{2}-1\right) \delta _{m+n,0}\,,  \notag
\\
\left[ \mathcal{J}_{m},\mathcal{G}_{r}\right] &=&\left( \frac{m}{2}-r\right)
\mathcal{G}_{m+r}\,,  \label{sbms3} \\
\left\{ \mathcal{G}_{r},\mathcal{G}_{s}\right\} &=&\mathcal{P}_{r+s}+\frac{%
c_{2}}{6}\,\left( r^{2}-\frac{1}{4}\right) \delta _{r+s,0}\,.  \notag
\end{eqnarray}%
The superalgebra obtained corresponds to the most generic super-BMS$_{3}$
algebra allowing two central charges \cite{BDMT, BDMT2}. In particular, the
central charges are associated to two terms in the CS Poincaré supergravity
action (\ref{csp}). Indeed, $c_{1}=12k\mu _{0}$ is related to the exotic CS
term, while $c_{2}=12k\mu _{1}$ is associated to the Einstein-Hilbert term.

Note that the Poincaré superalgebra is spanned by $\mathcal{J}_{0},\mathcal{J%
}_{1},\mathcal{J}_{-1},\mathcal{P}_{0},\mathcal{P}_{1},\mathcal{P}_{-1}$ and
$\mathcal{G}_{\frac{1}{2}},\mathcal{G}_{-\frac{1}{2}}$. This can be seen
explicitly considering the following change of basis:%
\begin{equation}
\begin{tabular}{lllll}
$\mathcal{J}_{-1}=-\sqrt{2}J_{0}$ & , & $\mathcal{J}_{1}=\sqrt{2}J_{1}$ & ,
& $\mathcal{J}_{0}=J_{2}\,,$ \\
$\mathcal{P}_{-1}=-\sqrt{2}P_{0}$ & , & $\mathcal{P}_{1}=\sqrt{2}P_{1}$ & ,
& $\mathcal{P}_{0}=P_{2}\,,$ \\
$\mathcal{G}_{-\frac{1}{2}}=\sqrt{2}Q_{+}$ & , & $\mathcal{G}_{\frac{1}{2}}=%
\sqrt{2}Q_{-}\,.$ &  &
\end{tabular}%
\end{equation}%
Then, the super-BMS$_{3}$ algebra (\ref{sbms3}) is the infinite-dimensional
lift of the three-dimensional Poincaré superalgebra.

Interestingly, we have used the same semigroup required to obtain the
super-Poincaré algebra from super-Lorentz. This show that the particular
procedure used to obtain a superalgebra can also be used to derive its
asymptotic symmetry. Although a similar result has been obtained at the
bosonic level for Poincaré and Maxwell CS\ gravity \cite{CCRS, CMMRSV}, this
is the first result at the supersymmetric level showing that the $S$%
-expansion method can be generalized to asymptotic symmetries.

\subsection{$\mathcal{N}=2$ super-BMS$_{3}$ algebra}

The extension to $\mathcal{N}=2$ super-BMS$_{3}$ algebra requires a more
subtle treatment. In particular, we will focus only in the $\mathcal{N}%
=\left( 2,0\right) $ case. Although we can extend our procedure to the $%
\left( 1,1\right) $ super-BMS$_{3}$ algebra, this would require to consider
a different starting superalgebra. In addition, the $\left( 1,1\right) $
super-BMS$_{3}$ algebra has no internal symmetry generators. As we shall
see, the $\left( 2,0\right) $ super-BMS$_{3}$ algebra obtained here
corresponds to a supersymmetric extension of the BMS$_{3}$ algebra endowed
with a $\mathfrak{\hat{u}}\left( 1\right) \times \mathfrak{\hat{u}}\left(
1\right) $ current algebra \cite{DR, BDR}. This is due to the fact that the $%
\left( 2,0\right) $ Poincaré superalgebra leading to a consistent CS\
supergravity action has a richer algebraic structure than the $\left(
1,1\right) $ case. Indeed, the $\left( 2,0\right) $ Poincaré superalgebra
includes an $\mathfrak{so}\left( 2\right) $ automorphism algebra \cite{HIPT}%
. Such interesting behavior is inherited to the asymptotic symmetry.
Recently, the authors of \cite{FMT} have shown than a democratic \cite{LM}
or ultra-relativistic IW contraction of the $\mathcal{N}=\left( 2,2\right) $
superconformal algebra reproduces the $\mathcal{N}=\left( 2,0\right) $
super-BMS$_{3}$ algebra. Here, we show that the $\left( 2,0\right) $
super-BMS$_{3}$ algebra can be alternatively obtained by expanding the $%
\mathcal{N}=2$ super-Virasoro algebra.

The generators of the $\mathcal{N}=2$ super-Virasoro algebra, which we shall
denote as $\mathfrak{svir}_{\left( 2\right) }$, satisfy the following
commutators%
\begin{eqnarray}
\left[ \ell _{m},\ell _{n}\right] &=&\left( m-n\right) \ell _{m+n}+\frac{c}{%
12}\,m\left( m^{2}-1\right) \delta _{m+n,0}\,,  \notag \\
\left[ \ell _{m},\mathcal{Q}_{r}^{i}\right] &=&\left( \frac{m}{2}-r\right)
\mathcal{Q}_{m+r}^{i}\,,  \notag \\
\left[ \ell _{m},\mathcal{R}_{n}\right] &=&-n\mathcal{R}_{m+n}\,,  \notag \\
\left[ \mathcal{R}_{m},\mathcal{R}_{n}\right] &=&\frac{c}{3}\,m\delta
_{m+n,0}\,,  \label{20sV} \\
\left[ \mathcal{Q}_{r}^{i},\mathcal{R}_{m}\right] &=&\epsilon ^{ij}\mathcal{Q%
}_{m+r}^{j}\,,  \notag \\
\left\{ \mathcal{Q}_{r}^{i},\mathcal{Q}_{s}^{j}\right\} &=&\delta ^{ij}\,%
\left[ \ell _{r+s}+\frac{c}{6}\,\left( r^{2}-\frac{1}{4}\right) \delta
_{r+s,0}\,\right] -2\epsilon ^{ij}\left( r-s\right) \mathcal{R}_{r+s}\,,
\notag
\end{eqnarray}%
where the central charge $c=12k$ is associated to the CS action (\ref{sLa}).
Such infinite-dimensional superalgebra differs from the $\mathcal{N}=1$
super-Virasoro one by the presence of an R-symmetry generator $\mathcal{R}%
_{m}$. Let us note that the $\mathcal{N}=2$ super-Virasoro algebra can be
decomposed in subspaces as%
\begin{equation}
\mathfrak{svir}_{\left( 2\right) }=V_{0}\oplus V_{1}\,,
\end{equation}%
where $V_{0}$ is the bosonic subspace spanned by the Virasoro generator $%
\ell _{m}$, the central charge $c$ and the R-symmetry generator $\mathcal{R}%
_{m}$. On the other hand, $V_{1}$ is the fermionic subspace. Such subspaces
satisfy a graded Lie algebra,%
\begin{eqnarray}
\left[ V_{0},V_{0}\right] &\subset &V_{0}\,,  \notag \\
\left[ V_{0},V_{1}\right] &\subset &V_{1}\,,  \label{SS} \\
\left[ V_{1},V_{1}\right] &\subset &V_{0}\,.  \notag
\end{eqnarray}

Let $S_{E}^{\left( 2\right) }=\left\{ \lambda _{0},\lambda _{1},\lambda
_{2},\lambda _{3}\right\} $ be the relevant semigroup whose elements satisfy
(\ref{ml}). The next step consists in considering a $0_{s}$-reduced resonant
subalgebra of $S_{E}^{\left( 2\right) }\times \mathfrak{svir}_{\left(
2\right) }$ following the definitions of \cite{Sexp}. In particular, the
following subset decomposition $S_{E}^{\left( 2\right) }=S_{0}\cup S_{1}$,
with%
\begin{eqnarray}
S_{0} &=&\left\{ \lambda _{0},\lambda _{2},\lambda _{3}\right\} \,,  \notag
\\
S_{1} &=&\left\{ \lambda _{1},\lambda _{3}\right\} \,,
\end{eqnarray}%
is resonant since it satisfies the same subspace structure (\ref{SS}). The
expanded infinite-dimensional superalgebra is generated by the set $\left\{
\mathcal{J}_{m},\mathcal{P}_{m},\mathcal{T}_{m},\mathcal{Z}_{m},\mathcal{G}%
_{r}^{i},c_{1},c_{2}\right\} $ whose generators and central charges are
related to the $\mathcal{N}=2$ super-Virasoro ones through:%
\begin{equation}
\begin{tabular}{ll}
$\mathcal{J}_{m}=\lambda _{0}\ell _{m}\,,$ & $c_{1}=\lambda _{0}c\,,$ \\
$\mathcal{P}_{m}=\lambda _{2}\ell _{m}\,,$ & $c_{2}=\lambda _{2}c\,,$ \\
$\mathcal{T}_{m}=\lambda _{0}\mathcal{R}_{m}\,,$ & $\mathcal{Z}_{m}=\lambda
_{2}\mathcal{R}_{m}\,,$ \\
$\mathcal{G}_{r}^{i}=\lambda _{1}\mathcal{Q}_{r}^{i}\,.$ &
\end{tabular}%
\end{equation}%
Using the (anti-)commutation relations of the $\mathcal{N}=2$ super-Virasoro
algebra along with the multiplication law of the semigroup (\ref{ml}), one
find that the (anti-)commutators of the expanded superalgebra read%
\begin{eqnarray}
\left[ \mathcal{J}_{m},\mathcal{J}_{n}\right] &=&\left( m-n\right) \mathcal{J%
}_{m+n}+\frac{c_{1}}{12}\,m\left( m^{2}-1\right) \delta _{m+n,0}\,,  \notag
\\
\left[ \mathcal{J}_{m},\mathcal{P}_{n}\right] &=&\left( m-n\right) \mathcal{P%
}_{m+n}+\frac{c_{2}}{12}\,m\left( m^{2}-1\right) \delta _{m+n,0}\,,  \notag
\\
\left[ \mathcal{J}_{m},\mathcal{T}_{n}\right] &=&-n\mathcal{T}%
_{m+n}\,,\qquad \left[ \mathcal{P}_{m},\mathcal{T}_{n}\right] =-n\mathcal{Z}%
_{m+n}\,,  \notag \\
\left[ \mathcal{J}_{m},\mathcal{Z}_{n}\right] &=&-n\mathcal{Z}_{m+n}\,,
\notag \\
\left[ \mathcal{T}_{m},\mathcal{T}_{n}\right] &=&\frac{c_{1}}{3}\,m\delta
_{m+n,0}\,,\qquad \left[ \mathcal{T}_{m},\mathcal{Z}_{n}\right] =\frac{c_{2}%
}{3}\,m\delta _{m+n,0}\,,  \label{20sBMS} \\
\left[ \mathcal{J}_{m},\mathcal{G}_{r}\right] &=&\left( \frac{m}{2}-r\right)
\mathcal{G}_{m+r}\,,  \notag \\
\left[ \mathcal{Q}_{r}^{i},\mathcal{T}_{m}\right] &=&\epsilon ^{ij}\mathcal{Q%
}_{m+r}^{j}\,,  \notag \\
\left\{ \mathcal{G}_{r}^{i},\mathcal{G}_{s}^{j}\right\} &=&\delta ^{ij}\left[
\mathcal{P}_{r+s}+\frac{c_{2}}{6}\,\left( r^{2}-\frac{1}{4}\right) \delta
_{r+s,0}\right] +2\epsilon ^{ij}\left( r-s\right) \mathcal{Z}_{r+s}\,.
\notag
\end{eqnarray}%
The infinite-dimensional superalgebra obtained corresponds to the $\mathcal{N%
}=\left( 2,0\right) $ super-BMS$_{3}$ algebra. Let us note that the
(anti-)commutator of the supercharges closes to a combination of $\mathcal{P}
$, central charge $c_{2}$ and $\mathcal{Z}$. The superalgebra obtained here
is found to be spanned by a supersymmetric extension of the enhanced
asymptotic symmetry algebra of 2+1 dimensional flat space which is given by
the BMS$_{3}$ algebra endowed with a $\mathfrak{\hat{u}}\left( 1\right)
\times \mathfrak{\hat{u}}\left( 1\right) $ current algebra \cite{DR, BDR}.
The explicit $\mathfrak{\hat{u}}\left( 1\right) $ current generators $%
\mathfrak{k}_{m}$ and $\mathfrak{\bar{k}}_{m}$ appear after the redefinitions%
\begin{equation}
\mathcal{T}_{m}=\mathfrak{k}_{m}-\mathfrak{\bar{k}}_{-m}\,,\qquad \mathcal{Z}%
_{m}=\epsilon \left( \mathfrak{k}_{m}+\mathfrak{\bar{k}}_{-m}\right) \,.
\end{equation}%
In particular, (\ref{20sBMS}) is recovered in the limit $\epsilon
\rightarrow 0$.

Let us note that the $\left( 2,0\right) $ super-Poincaré algebra is spanned
by $\left\{ \mathcal{J}_{m},\mathcal{P}_{n},\mathcal{G}_{r}^{i},\mathcal{T}%
_{0},\mathcal{Z}_{0}\right\} $ with $m,n=0,\pm 1$ and $r=\pm \frac{1}{2}$.
In fact, the (anti)-commutation relations (\ref{sp1})-(\ref{sp2}) for $%
\mathcal{N}=\left( 2,0\right) $ appear explicitly after the redefinitions%
\begin{equation}
\begin{tabular}{lllll}
$\mathcal{J}_{-1}=-\sqrt{2}J_{0}$ & , & $\mathcal{J}_{1}=\sqrt{2}J_{1}$ & ,
& $\mathcal{J}_{0}=J_{2}\,,$ \\
$\mathcal{P}_{-1}=-\sqrt{2}P_{0}$ & , & $\mathcal{P}_{1}=\sqrt{2}P_{1}$ & ,
& $\mathcal{P}_{0}=P_{2}\,,$ \\
$\mathcal{G}_{-\frac{1}{2}}^{i}=\sqrt{2}Q_{+}^{i}$ & , & $\mathcal{G}_{\frac{%
1}{2}}^{i}=\sqrt{2}Q_{-}^{i}\,$ & , &  \\
$\mathcal{T}_{0}=-T$ & , & $\mathcal{Z}_{0}=-Z\,.$ &  &
\end{tabular}
\label{redef}
\end{equation}%
Then, the $\mathcal{N}=2\,\ $super-BMS$_{3}$ algebra (\ref{20sBMS}) is the
infinite-dimensional lift of the three-dimensional $\left( 2,0\right) $
Poincaré superalgebra endowed with an automorphism generator $T$ and central
charge $Z$. Let us note that the $\mathcal{N}=2$ super-BMS$_{3}$ given by (\ref{20sBMS}) can be alternatively derived from the $\mathcal{N}=4$ one appearing in \cite{BJLMN}. Indeed the (anti-)commutation relations (\ref{20sBMS}) can be easily obtained from $\mathcal{N}=4$ super BMS$_{3}$ of \cite{BJLMN} after setting some fermionic generators to zero. An inequivalent $\mathcal{N}=2$ super-BMS$_{3}$ algebra can also
be obtained considering a "despotic" \cite{LM} contraction of the $\mathcal{N%
}=\left( 2,2\right) $ superconformal algebra.

\subsection{$\mathcal{N}=4$ super-BMS$_{3}~$algebra}

For completeness, we extend our construction to the $\mathcal{N}=4$ case.
Obtaining a $\mathcal{N}=4$ super-BMS$_{3}$ algebra, following our
procedure, requires to $S$-expand a $\mathcal{N}=4$ super-Virasoro algebra.
Here, we shall focus our attention to the derivation of a super-BMS$_{3}$
algebra whose finite subalgebra is not only the $\mathcal{N}=4$ super-Poincar%
é algebra but also the internal algebra.

The $\mathcal{N}=4$ super-Virasoro algebra reads as \cite{Ito}:
\begin{eqnarray}
\left[ \ell _{m},\ell _{n}\right] &=&\left( m-n\right) \ell _{m+n}+\frac{c}{%
12}\,m\left( m^{2}-1\right) \delta _{m+n,0}\,,  \notag \\
\left[ \ell _{m},\mathcal{Q}_{r}^{i,\pm }\right] &=&\left( \frac{m}{2}%
-r\right) \mathcal{Q}_{m+r}^{i,\pm }\,,  \notag \\
\left[ \ell _{m},\mathcal{R}_{n}^{a}\right] &=&-n\mathcal{R}_{m+n}^{a}\,,
\notag \\
\left[ \mathcal{R}_{m}^{a},\mathcal{R}_{n}^{b}\right] &=&i\epsilon ^{abc}%
\mathcal{R}_{m+n}^{c}+\frac{c}{12}\,m\delta ^{ab}\delta _{m+n,0}\,,
\label{22sV} \\
\left[ \mathcal{R}_{m}^{a},\mathcal{Q}_{r}^{i,+}\right] &=&\,-\frac{1}{2}%
\left( \sigma ^{a}\right) _{j}^{i}\mathcal{Q}_{m+r}^{j,+}\,,\qquad \left[
\mathcal{R}_{m}^{a},\mathcal{Q}_{r}^{i,-}\right] =\,\frac{1}{2}\left( \bar{%
\sigma}^{a}\right) _{j}^{i}\mathcal{Q}_{m+r}^{j,-}\,,  \notag \\
\left\{ \mathcal{Q}_{r}^{i,+},\mathcal{Q}_{s}^{j,-}\right\} &=&\delta ^{ij}\,%
\left[ \ell _{r+s}+\frac{c}{6}\,\left( r^{2}-\frac{1}{4}\right) \delta
_{r+s,0}\,\right] -\left( r-s\right) \left( \sigma ^{a}\right) _{ij}\mathcal{%
R}_{r+s}^{a}\,,  \notag
\end{eqnarray}%
where $i,j=1,2$; $a,b,c=1,2,3$ and $\bar{\sigma}_{ij}^{a}=\sigma _{ji}^{a}$
are the Pauli matrices. It is important to emphasize that the superalgebra (%
\ref{22sV}) considered here contains only one set of Virasoro generators $%
\ell _{m}$. In particular, the generators $\left\{ \mathcal{\ell }_{m},%
\mathcal{Q}_{r}^{i,\pm },\mathcal{R}_{0}^{a}\right\} $ with $m=0,\pm 1$ and $%
r=\pm \frac{1}{2}$ satisfy a finite subalgebra which corresponds to an $%
\mathcal{N}=4$ super-Lorentz.

Let $S_{E}^{\left( 2\right) }=\left\{ \lambda _{0},\lambda _{1},\lambda
_{2},\lambda _{3}\right\} $ be the relevant semigroup with the
multiplication law given by (\ref{ml}). Considering the resonant
decomposition (\ref{sd2}) and applying a $0_{s}$-reduction resonant $%
S_{E}^{\left( 2\right) }$-expansion of the $\mathcal{N}=4$ super Virasoro
algebra we find a bigger infinite-dimensional superalgebra spanned by%
\begin{equation}
\left\{ \mathcal{J}_{m},\,\mathcal{P}_{m},\,\mathcal{T}_{m}^{a},\,\mathcal{Z}%
_{m}^{a},\,\mathcal{G}_{r}^{i,\pm },\,c_{1},\,c_{2}\right\} \,.
\end{equation}%
Such generators are related to the super-Virasoro ones through the semigroup
elements as%
\begin{equation}
\begin{tabular}{ll}
$\mathcal{J}_{m}=\lambda _{0}\ell _{m}\,,$ & $c_{1}=\lambda _{0}c\,,$ \\
$\mathcal{P}_{m}=\lambda _{2}\ell _{m}\,,$ & $c_{2}=\lambda _{2}c\,,$ \\
$\mathcal{T}_{m}^{a}=\lambda _{0}\mathcal{R}_{m}^{a}\,,$ & $\mathcal{Z}%
_{m}^{a}=\lambda _{2}\mathcal{R}_{m}^{a}\,,$ \\
$\mathcal{G}_{r}^{i,\pm }=\lambda _{1}\mathcal{Q}_{r}^{i,\pm }\,.$ &
\end{tabular}%
\end{equation}%
Using the multiplication law of the semigroup (\ref{ml}) and the
(anti-)commutation relations of the $\mathcal{N}=4$ super-Virasoro algebra,
one find that the expanded generators satisfy the following non-vanishing
(anti-)commutators%
\begin{eqnarray}
\left[ \mathcal{J}_{m},\mathcal{J}_{n}\right] &=&\left( m-n\right) \mathcal{J%
}_{m+n}+\frac{c_{1}}{12}\,m\left( m^{2}-1\right) \delta _{m+n,0}\,,  \notag
\\
\left[ \mathcal{J}_{m},\mathcal{P}_{n}\right] &=&\left( m-n\right) \mathcal{P%
}_{m+n}+\frac{c_{2}}{12}\,m\left( m^{2}-1\right) \delta _{m+n,0}\,,  \notag
\\
\left[ \mathcal{J}_{m},\mathcal{T}_{n}^{a}\right] &=&-n\mathcal{T}%
_{m+n}^{a}\,,\qquad \left[ \mathcal{P}_{m},\mathcal{T}_{n}^{a}\right] =-n%
\mathcal{Z}_{m+n}^{a}\,,  \notag \\
\left[ \mathcal{J}_{m},\mathcal{Z}_{n}^{a}\right] &=&-n\mathcal{Z}%
_{m+n}^{a}\,,  \notag \\
\left[ \mathcal{T}_{m}^{a},\mathcal{T}_{n}^{b}\right] &=&i\epsilon ^{abc}%
\mathcal{T}_{m+n}^{c}+\frac{c_{1}}{12}\,m\delta ^{ab}\delta _{m+n,0}\,, \\
\left[ \mathcal{T}_{m}^{a},\mathcal{Z}_{n}^{b}\right] &=&i\epsilon ^{abc}%
\mathcal{Z}_{m+n}^{c}+\frac{c_{2}}{12}\,m\delta ^{ab}\delta _{m+n,0}\,,
\notag \\
\left[ \mathcal{J}_{m},\mathcal{G}_{r}^{i,\pm }\right] &=&\left( \frac{m}{2}%
-r\right) \mathcal{G}_{m+r}^{i,\pm }\,,  \notag \\
\left[ \mathcal{T}_{m}^{a},\mathcal{G}_{r}^{i,+}\right] &=&\,-\frac{1}{2}%
\left( \sigma ^{a}\right) _{j}^{i}\mathcal{G}_{m+r}^{j,+}\,,\qquad \left[
\mathcal{T}_{m}^{a},\mathcal{G}_{r}^{i,-}\right] =\,\frac{1}{2}\left( \bar{%
\sigma}^{a}\right) _{j}^{i}\mathcal{G}_{m+r}^{j,-}\,,  \notag \\
\left\{ \mathcal{G}_{r}^{i,+},\mathcal{G}_{s}^{j,-}\right\} &=&\delta ^{ij}\,%
\left[ \mathcal{P}_{r+s}+\frac{c_{2}}{6}\,\left( r^{2}-\frac{1}{4}\right)
\delta _{r+s,0}\,\right] -\left( r-s\right) \left( \sigma ^{a}\right) _{ij}%
\mathcal{Z}_{r+s}^{a}\,.  \notag
\end{eqnarray}%
The $S$-expanded algebra corresponds to a $\mathcal{N}=4$ supersymmetric
extension of the BMS$_{3}$ algebra. In particular, the (anti-)commutators of
such infinite-dimensional symmetry close into a linear combination of $%
\mathcal{P}$ and their respective central charges and $\mathcal{Z}^{a}$
generators. Interestingly, one can show that $\mathfrak{su}\left( 2\right) $
current generators $\mathfrak{k}_{m}^{a}$ and $\mathfrak{\bar{k}}_{m}^{a}$
appear after the redefinitions%
\begin{equation}
\mathcal{T}_{m}^{a}=\lim_{\epsilon \rightarrow 0}\left( \mathfrak{k}_{m}^{a}-%
\mathfrak{\bar{k}}_{-m}^{a}\right) \,,\qquad \mathcal{Z}_{m}=\lim_{\epsilon
\rightarrow 0}\epsilon \left( \mathfrak{k}_{m}^{a}+\mathfrak{\bar{k}}%
_{-m}^{a}\right) \,.
\end{equation}%
Note that the $\mathcal{N}=4$ super-BMS$_{3}$ obtained here contains note
only the $\mathcal{N}=4$ super-Poincaré algebra as the finite subalgebra,
but also the internal algebra generated by the $\mathcal{T}_{m}^{a}$
generators. In fact, the set $\left\{ \mathcal{J}_{m},\mathcal{P}_{n},%
\mathcal{G}_{r}^{i,\pm },\mathcal{T}_{0}^{a},\mathcal{Z}_{0}^{a}\right\} $,
with $m,n=0,\pm 1$ and $r=\pm \frac{1}{2}$, generates the $\mathcal{N}=4$
super-Poincaré algebra endowed with an internal algebra. On the other hand,
the central charges $c_{1}=12k\mu _{0}$ and $c_{2}=12k\mu _{1}$ are related
to the CS level (\ref{csp}).

Alternative approaches have been considered in \cite{BJLMN, BLN} where alternative $\mathcal{N}=4$ super-BMS$_{3}$ algebras have been presented. In
particular, the supersymmetric extension of the BMS$_{3}$ algebra\ of \cite%
{BJLMN, BLN} contains R-symmetry generators $\mathcal{R}_{m}$ which come
from two copies of the $\mathcal{N}=2$ super Virasoro algebra. Since for our
construction we are considering only one $\mathcal{N}=4$ super-Virasoro
algebra, the final $\mathcal{N}=4$ super-BMS$_{3}$ contains the presence of
internal symmetry generators $\mathcal{T}_{m}^{a}$ instead of $\mathcal{R}%
_{m}$ generators. As was noticed in \cite{HIPT}, the presence of
automorphism generators are essential in order to define a non-degenerate
invariant inner product which allows the formulation of a CS supergravity
action.\

\section{Conclusions}

In this paper we have presented a novel approach to obtain the $\mathcal{N}$%
-extended Poincaré supergravity and its respective asymptotic symmetry: the $%
\mathcal{N}$-extended super-BMS$_{3}$ algebra. This alternative approach is
based on the semigroup expansion method. In particular, we have shown that
the $\mathcal{N}$-extended super-Poincaré algebra with both central and
automorphism generators appears by expanding the super Lorentz with a
particular semigroup $S_{E}^{\left( 2\right) }$. Interestingly,\
three-dimensional flat supergravity appears naturally from an exotic
supersymmetric theory based only on the spin-connection. This peculiarity
manifests itself only in three spacetime dimensions since there is the same
number of Lorentz and boost generators allowing us to identify the expanding
Lorentz fields as vielbein.

Remarkably, we have extended our results to infinite-dimensional algebras to
get the $\mathcal{N}$-extended super-BMS$_{3}$ algebra for $\mathcal{N}=(1,2,4)$. It is worth it to
mention that such supersymmetric extensions of the BMS$_{3}$ symmetry are
obtained by expanding one Virasoro superalgebra using the same finite
semigroup $S_{E}^{\left( 2\right) }$ as we can see in the following diagram:%
\begin{equation*}
\begin{tabular}{ccc}
\cline{1-1}\cline{3-3}
\multicolumn{1}{|c}{$%
\begin{array}{c}
\mathcal{N}\text{-extended} \\
\text{super Lorentz}%
\end{array}%
$} & \multicolumn{1}{|c}{$\overset{\text{infinite-dimensional lift}}{%
\longrightarrow }$} & \multicolumn{1}{|c|}{$%
\begin{array}{c}
\mathcal{N}\text{-extended} \\
\text{super Virasoro}%
\end{array}%
$} \\ \cline{1-1}\cline{3-3}
&  &  \\
$\ \ \downarrow $ $S_{E}^{\left( 2\right) }$ &  & $\ \ \downarrow $ $%
S_{E}^{\left( 2\right) }$ \\
&  &  \\ \cline{1-1}\cline{3-3}
\multicolumn{1}{|c}{$%
\begin{array}{c}
\mathcal{N}\text{-extended} \\
\text{super Poincaré}%
\end{array}%
$} & \multicolumn{1}{|c}{$\overset{\text{infinite-dimensional lift}}{%
\longrightarrow }$} & \multicolumn{1}{|c|}{$%
\begin{array}{c}
\mathcal{N}\text{-extended} \\
\text{super-BMS}_{3}%
\end{array}%
$} \\ \cline{1-1}\cline{3-3}
\end{tabular}%
\end{equation*}

Of particular interest are the $\mathcal{N}>1$ super-BMS$_{3}$ algebras
obtained here since they are not only centrally extended but also endowed
with internal symmetry algebra.
Interestingly, we have shown that $\mathcal{N}$-extended Poincaré
superalgebras with both central and automorphism charges \cite{HIPT} appear
as finite subalgebras of the $\mathcal{N}$-extended super-BMS$_{3}$
constructed here. It is interesting to note that the $\mathcal{N}=2$ super-BMS$_{3}$ presented here can be easily obtained from the $\mathcal{N}=4$ one presented in \cite{BJLMN} after setting some fermionic generators to zero.

Our results are not only a supersymmetric generalization of those presented
in \cite{CCRS} but could also be extended to other infinite-dimensional
supersymmetries. It has been recently introduced in \cite{CCRS}, using the $%
S $-expansion procedure, an enlarged and deformed BMS$_{3}$ algebra which
can be seen as a infinite-dimensional lift of the Maxwell algebra.
Interestingly, this new infinite-dimensional algebra results to be the
corresponding asymptotic symmetry of the three-dimensional CS gravity for
the Maxwell algebra \cite{CMMRSV}. Subsequently, in \cite{CMRSV}, a semi-simple enlargement of the BMS$_3$ algebra has been presented as the asymptotic symmetry of the AdS-Lorentz CS gravity. Then, motivated by these recent results,
it would be interesting to explore the supersymmetric extension of these
deformed and enlarged BMS$_{3}$ algebras using the same methodology
considered here [work in progress]. One could expect that such
supersymmetrization is the corresponding asymptotic symmetry of a CS
supergravity \cite{CPR} in three spacetime dimensions for the Maxwell and AdS-Lorentz superalgebra.

Another natural generalization of our results is the extension of our
procedure to the complete family of super Maxwell like algebras \cite{CR1, CDMR}.
It has been pointed out in \cite{CCRS} that the BMS$_{3}$ and deformed BMS$%
_{3}$ belong to a larger family of infinite-dimensional symmetry. One could
expect to obtain the complete family of infinite-dimensional $\mathcal{N}$%
-extended superalgebras in which the super-BMS$_{3}$ is a particular case.

\section{Acknowledgment}

This work was supported by the Chilean FONDECYT Projects N$^{\circ }$3170437
(P.C.), N$^{\circ }$3170438 (E.R.). R.C. appreciates the support of the
Special Financing of Academic Activities of the Universidad Católica de la
Santísima Concepción, Chile. R.C. and O.F. would like to thank to the Direcci%
ón de Investigación and Vice-rectoría de Investigación of the Universidad Cat%
ólica de la Santísima Concepción, Chile, for their constant support.

\end{document}